\documentclass[aps,amssymb,amsmath,twocolumn,pra,superscriptaddress]{revtex4}
\usepackage[pdftex]{graphicx}
\usepackage{dcolumn}
\usepackage{bm}
\usepackage{hyperref}
\hypersetup{colorlinks=true}
\usepackage{color}
\usepackage[up]{subfigure}

\newcommand{\be}{\begin{equation}}
\newcommand{\ee}{\end{equation}}
\newcommand{\bc}{\begin{center}}
\newcommand{\ec}{\end{center}}
\newcommand{\bea}{\begin{eqnarray}}
\newcommand{\eea}{\end{eqnarray}}
\newcommand{\ba}{\begin{array}}
\newcommand{\ea}{\end{array}}

\begin{document}
\title{Enhancement of Geometric Phase by Frustration of Decoherence: \\
A Parrondo like Effect}

\author{Subhashish \surname{Banerjee}}
\email{subhashish@iitj.ac.in}
\affiliation{Indian Institute of Technology, Rajasthan, Jodhpur 342011, India}

\author{C. M.  \surname{Chandrashekar}}
\email{c.madaiah@oist.jp}
\affiliation{Ultracold Quantum Gases, Physics Department, University College Cork, Cork, Ireland}
\affiliation{Quantum Systems Unit, Okinawa Institute of Science and Technology, Okinawa, Japan}

\author{Arun K. \surname{Pati}}
\email{akpati@hri.res.in}
\affiliation{Harish-Chandra Research Institute, Chhatnag Road, Jhunsi, Allahabad-211019, India}

\begin{abstract}
Geometric phase plays an important role in evolution of pure or mixed quantum states. However, when a system undergoes decoherence the development of 
geometric phase may be inhibited. Here, we show that when a quantum system interacts with two competing environments there can be enhancement of
geometric phase. This effect is akin to Parrondo like effect on the geometric phase which results from quantum frustration of decoherence. Our result
suggests that the mechanism of two competing decoherence can be useful in fault-tolerant holonomic quantum computation.
\end{abstract}


\maketitle
\section{Introduction}
\label{intro}

Geometric phase (GP) is a consequence of the holonomy of the path traced by a quantum system in its Hilbert space, thereby highlighting its connection to
the intrinsic curvature of the space\,\cite{simon}. 
Even though its classical foundation was laid by Pancharatnam\,\cite{panch}, in dealing with questions related to the
characterization  of interference of classical light in distinct states of polarization, its quantum counterpart was discovered much later by Berry\,\cite{berry} for cyclic
adiabatic evolution.  This was subsequently generalized to non-adiabatic\,\cite{aharonov}  and non-cyclic evolutions\,\cite{samuel}. 
Later, a quantum kinematic approach was provided for the GP\,\cite{mukunda}, and a generalized gauge potential for the most general quantum evolution was introduced\,\cite{akp}.
The concept of geometric phase is not limited to pure state quantum evolution, but does appear for mixed states \cite{uhlman,akp1,tong} also. An experimentally measurable geometric
phase for mixed states under unitary evolution was first introduced in Ref.\,\cite{akp1} and then generalized to nonunitary evolutions\,\cite{tong}.
Since the geometric phase depends on the evolution path and not on the detailed dynamics, thereby suggesting an inherent fault tolerance\,\cite{duan}, it can be a useful 
resource for quantum computation. Using Nunlear Magnetic Resonance\,\cite{SMP+88} and atom interferometry\,\cite{WGS99} pure and mixed state geometric phases have been realized experimentally.  
There have been various other proposals to observe GP in a coupled two-mode Bose-Einstein condensate\,\cite{fpb02}, Bose-Einstein Josephson junction\,\cite{be}, and superconducting nanostructure\,\cite{super}, in all of which 
it is imperative to consider the effect of the ambient environment on the system of interest\,\cite{env}.  
Further, the importance of GP in quantum computation can be gauged from its proposal and experimental realization in
ion traps \cite{duan, it}, cavity quantum electrodynamics \cite{caveqed}, non-Abelian GP in atomic ensembles \cite{ae} and in quantum dots \cite{qd}.
Recently, there have been attempts to connect GP with 
quantum correlations, in particular entanglement\,\cite{sarandy}, in a variety of quantum systems.  

   The above reasons bring to focus the need to have an understanding of the impact of the environment on the study and practical implementation of GP. 
In fact, the effect of measurement on the GP was first investigated in Ref.\,\cite{akp2} and it was shown that in the limit of continuous observation 
the GP can be suppressed. For mixed states it was shown that the Uhlmann phase also decreases under isotropic decoherence\,\cite{erik}.
Open quantum systems make up the systematic study of the influence of the environment, alternatively called the reservoir or bath, on the evolution of the system of interest.
The basic idea is that one follows the evolution of the system of interest by tracing out the 
environmental degrees of freedom, resulting in a non-unitary evolution. Decoherence and dissipation are a natural consequence of this. 
Open quantum systems can be broadly classified into two categories,
one that involves decoherence without dissipation\,\cite{bg07, srigp} and the other where dissipation occurs along with decoherence\,\cite{sq, srigp}.  Experiments with trapped atoms
have been performed where both pure decoherence as well as a dissipative type of evolution have been generated by coupling the atomic system to appropriate engineered reservoirs
\,\cite{turchette}.   A practical implementation of GP would involve, for example, a qubit interacting with its environment, resulting in its inhibition. This calls for the need to have
settings where the inhibition of GP, due to the ubiquitous environment, could be arrested.  Quantum frustration of decoherence (QFD), as demonstrated in this paper, would be a potential 
candidate for achieving this. 

QFD is the term ascribed to the general phenomena when a quantum system coupled to two independent environments by canonically conjugate operators results
in an enhancement of quantum fluctuations, that is, decoherence gets suppressed\,\cite{novais}.  The reason for this is attributable to the non-commuting nature of the 
conjugate coupling operators that prevents the selection of an appropriate pointer basis to which the quantum system could settle down. It has been studied in various
guises, such as an extension of the dissipative two-level system problem\,\cite{novais}, where the two non-commuting spin operators of the central spin system were coupled to 
independent harmonic oscillator baths, or a harmonic oscillator, modeling a large spin impurity in a ferromagnet, coupled to two independent oscillator baths via its position
and momentum operators\,\cite{kohler}. In each case, irrespective of the system of interest or the coupling operators, QFD was 
observed.  Another scenario where this has been put to use is in quantum error correction\,\cite{novais1}. These considerations
were extended to the case of spin baths\,\cite{rao}, present, for example, in the case of quantum dots, with similar results. These motivate us to study GP in the presence of QFD. 
Interestingly, this could be also thought of as an example of Parrondo's paradox involving two games which when played individually
lead to a loosing expectation, but when played in an alternative order produce a winning expectation\,\cite{abbott1,abbott2}. The underlying reason behind the surprising aspect 
of Parrondo's game is the breaking of an inherent symmetry in the problem. This feature is also shared by quantum frustration models where the symmetry in the decay channel, 
were only one bath present, is broken by the presence of coupling to two independent baths by non-commuting operators. Here, we take up a simple model of a frustrated open
quantum system and explicitly show the enhancement of GP. This highlights the role of quantum frustrated decoherence leading to a Parrondo like effect on the geometric phase.

The paper is organized as follows. In Section\,\ref{model} we introduce the model of a spin interacting with two independent spin baths to show the influence of QFD on GP. In Section\,\ref{GPanalysis} we present the explicit solution and analysis 
for the GP of frustrated spin system. In Section\,\ref{analogy} we present the analogy of the GP dynamics with Parrondo games and conclude in section\,\ref{conc}. 
 
\section{Model}
\label{model}
 We study the influence of QFD on GP by taking up a simple model involving a central spin, or a qubit which would be our system of interest,
interacting with two independent spin baths via two non-commuting spin operators
\begin{eqnarray}
\label{eq:1}
H &=& H_S + H_{SR} \nonumber\\
&=& \omega \frac{\sigma_z}{2} + \alpha_1 \frac{\sigma_x}{2} \otimes  \Sigma_{k=1}^N I_x^k + \alpha_2 \frac{\sigma_y}{2} \otimes \Sigma_{l=1}^N J_y^l,
\end{eqnarray}
where $H_S $ is the system (single qubit) Hamiltonian and  $H_{SR}$
is the system-reservoir interaction Hamiltonian.  Here $\sigma_i$, $i = x, y, z$ are the three Pauli matrices for the central spin, and $I_x^k$ and $J_y^l$ are the 
bath spin operators. Also,  $\alpha_1$, $\alpha_2$ are the two spin-bath coupling constants and  $\omega$ comes from the basic system Hamiltonian, representing the initial magnetic field. 
The bath dynamics itself is not considered. This serves two purposes;  it allows for an analytical treatment of the model and at the same time captures its essence, since in solid state spin systems with dominant spin-environment interactions, such as quantum dots where 
such a model could be envisaged, the internal bath dynamics composed of nuclear spins would be very slow compared to the central electronic spin \cite{loss}. 

Assume an uncorrelated system-reservoir initial state with the central spin in 
\bea
\label{singlequbit}
\rho_S(0) &=& \cos^2 \left ( \frac{\theta}{2} \right ) |\downarrow \rangle \langle \downarrow| + \sin^2 \left (\frac{\theta}{2} \right ) |\uparrow \rangle \langle \uparrow|  \nonumber\\
&+& \frac{i}{2} \sin(\theta) e^{i\phi}\left[|\uparrow \rangle \langle \downarrow| -e^{-i 2 \phi} |\downarrow \rangle \langle \uparrow| \right].
\eea
Equation\,(\ref{singlequbit}) is the most general single qubit density matrix where $\theta \in \{0, \pi\}$ and $\phi \in \{0, 2\pi\}$ are the polar and azimuthal angles, respectively. 
The full form of the initial density matrix with an unpolarized initial bath state is 
$\rho_{SR} (0) = \frac{1}{2^{2N}} \rho_S (0) \otimes \mathcal{I}_{2^N} \otimes \mathcal{I}_{2^N}$, where $N$ is the total number of spins present in each bath. 
Under the interaction Hamiltonian, the total state evolves as $\rho_{SR} (0) \rightarrow \rho_{SR} (t) = \exp[-i(H_S + H_{SR})t] \rho_{SR} (0)\exp[i(H_S + H_{SR})t]$. 
After interaction, the reduced state of the spin is given by $\rho_S(t) = {\rm Tr}_R[\rho_{SR}(t)]$.
The Bloch vector representation of a spin-$\frac{1}{2}$ particle,  
which is the central spin here, is
\bea
  \rho_S(t) = \frac{1}{2}\begin{bmatrix}
 1+ \langle \sigma_z (t) \rangle  & & \langle \sigma_x (t) \rangle - i\langle \sigma_y (t) \rangle \\
\langle \sigma_x (t) \rangle + i\langle \sigma_y (t) \rangle & &  1- \langle \sigma_z (t) \rangle  \end{bmatrix}, \label{reduceddm}
\eea
where $\langle \sigma_i (t) \rangle = \sum_{m_1,m_2 = -N/2}^{N/2}   \zeta_{m_1} \zeta_{m_2} {\rm Tr} \left(\rho_{m_1,m_2}(0) \sigma_i(t)\right)$ 
and $m_{1}$, $m_{2}$ label the eigenvalues of bath spin operators and range from $-\frac{N}{2}$ to $\frac{N}{2}$.  
The average polarizations of the central spin come out to be
\bea
\langle \sigma_z (t) \rangle &=& \frac{-1}{2^{2N + 1}}\sum_{m_1,m_2 = -N/2}^{N/2}  \zeta_{m_1} \zeta_{m_2} \Bigl ( \cos(\Gamma_{m_1 m_2}t) \cos(\theta)   \nonumber\\
&+& \left. \frac{\sin(\Gamma_{m_1 m_2}t)}{\Gamma_{m_1 m_2}} \sin(\theta) \left[m_1 \alpha_1 \cos(\phi) - m_2 \alpha_2 \sin(\phi)\right] \right. \nonumber\\ &+& \omega
\left(m_1 \alpha_1 \sin(\theta) \sin(\phi) + m_2 \alpha_2 \sin(\theta) \cos(\phi) \right. \nonumber\\ &+& \left.\omega \cos(\theta)\right)
\frac{(1-\cos(\Gamma_{m_1 m_2}t))}{\Gamma_{m_1m_2}^2}\Bigr ), \label{zpolarized}
\eea
\bea
\langle \sigma_x (t) \rangle &=& \frac{-1}{2^{2N + 1}}\sum_{m_1,m_2 = -N/2}^{N/2}  \zeta_{m_1} \zeta_{m_2} \Bigl( \cos(\Gamma_{m_1 m_2}t)  \nonumber\\ &&  \sin(\theta) \sin(\phi) -
\frac{\sin(\Gamma_{m_1 m_2}t)}{\Gamma_{m_1 m_2}} \nonumber\\ &\times&  \left[\omega \sin(\theta)  \cos(\phi) - m_2 \alpha_2 \cos(\theta)\right]  \nonumber\\ &+&   m_1 \alpha_1 
\left(m_1 \alpha_1 \sin(\theta) \sin(\phi) + m_2 \alpha_2 \sin(\theta) \cos(\phi) \right. \nonumber\\ &+& \left. \omega \cos(\theta)\right)
\frac{(1 - \cos(\Gamma_{m_1 m_2}t))}{\Gamma_{m_1 m_2}^2}\Bigr), \label{xpolarized} 
\eea
and
\bea
\langle \sigma_y (t) \rangle &=& \frac{-1}{2^{2N + 1}}\sum_{m_1,m_2 = -N/2}^{N/2}  \zeta_{m_1} \zeta_{m_2} \Bigl( \cos(\Gamma_{m_1 m_2}t)  \nonumber\\ &&  \sin(\theta) \cos(\phi) -
\frac{\sin(\Gamma_{m_1 m_2}t)}{\Gamma_{m_1 m_2}}  \nonumber\\ &\times& \left[-\omega \sin(\theta)  \sin(\phi) + m_1 \alpha_1 \cos(\theta)\right]  \nonumber\\ &+&  m_2 \alpha_2 
\left(m_1 \alpha_1 \sin(\theta) \sin(\phi) + m_2 \alpha_2 \sin(\theta) \cos(\phi) \right.  \nonumber\\ &+&  \left. \omega \cos(\theta)\right)
\frac{(1 - \cos(\Gamma_{m_1 m_2}t))}{\Gamma_{m_1 m_2}^2}\Bigr ). \label{ypolarized} 
\eea
Here $\zeta_m = \frac{N!}{(N/2 - m)! (N/2 + m)!}$
and $\Gamma_{m_1 m_2}=\sqrt{\omega^2 + \alpha_1^2 m_1^2 + \alpha_2^2 m_2^2}$.

The vector $v(t) = \rm{Tr}\Big[\rho_S (t) \sigma(0)\Big]$ is called the Bloch vector of the system.
For pure states $|v(t)| = 1$ while for mixed states, $|v(t)| < 1$, that is, the Bloch vector penetrates into the Bloch sphere.

\section{GP of frustrated spin system: Explicit solution and analysis}
\label{GPanalysis}

A general mixed state density matrix $\rho(t) = \sum_k \lambda(k) |\phi_k(t)\rangle \langle \phi_k(t)| $ is subject to purification, by the introduction of an ancilla, as
\begin{equation}
|\Psi(t)\rangle =\sum_k \sqrt{\lambda(k)} |\phi_k(t)\rangle \otimes |a_k\rangle ; t \in [0,\tau],
\end{equation}
where $\lambda(k)$, $|\phi_k(t)\rangle$ are the eigenvalues and eigenvectors of the reduced density matrix $\rho(t)$ under consideration, respectively, 
and $|a_k\rangle$ represent the ancilla. The Pancharatnam relative phase, $\alpha (t)= arg(\langle \Psi(0)| \Psi(t)\rangle)$ reduces to the GP when 
the parallel transport condition, $\langle \phi_k(t) | d/dt |\phi_k(t)\rangle =0, k=1...P$ corresponding to the $P$ eigenstates, is satisfied. The GP for the 
mixed state, $\rho_S (t)$ [Eq.\,(\ref{reduceddm})], satisfying the parallel transport conditions assumes the form 
\bea
\gamma_g(\tau)=arg\Bigl[\sum_k\sqrt{\lambda_k(\tau)\lambda_k(0)}\langle \phi_k(0)|\phi_k(\tau)\rangle \nonumber \\
\times e^{-\int_0^{\tau} \langle \phi_k(t')|
\dot \phi_k(t')\rangle dt'} \Bigr] \label{GP}
\eea
where $\lambda_k(\tau)$ are the eigenvalues and $\phi_k(\tau)$ are the corresponding eigenvectors of the reduced density matrix $\rho_S (\tau)$ [Eq.\, (\ref{reduceddm})]. 

Equation\,(\ref{GP}) can be shown to be
\begin{eqnarray}
\gamma_g(\tau) &=& \arg\Bigl[\left\{\frac{1}{2} \left( 1 + 
\sqrt{A^2 (\tau) + 4 R^2(\tau)}\right)\right\}^{\frac{1}{2}} \nonumber\\
&\times& \left\{\cos\left(\frac{\theta_0}{2}\right)
\sin\left(\frac{\theta_{\tau}}{2}\right) \right. \nonumber \\
&+ & \left. e^{i (\chi(\tau) - \chi(0))}
\sin\left(\frac{\theta_0}{2}\right)
\cos\left(\frac{\theta_{\tau}}{2}\right)\right\} \nonumber \\ 
&\times& e^{-i \int_0^{\tau} dt (\dot{\chi}(t)) 
\cos^2\left ( \frac{\theta_t}{2}\right )}\Bigr ]. \label{4v}
\end{eqnarray}
 Here $A = \langle \sigma_z (t) \rangle$, $R = \frac{1}{2}\sqrt{\langle \sigma_x (t) \rangle)^2 + \langle \sigma_y (t) \rangle)^2}$ and $\tan (\chi(t)) = 
\frac{\langle \sigma_y (t) \rangle}{\langle \sigma_x (t) \rangle}$. Also,
$\sin\left(\frac{\theta_t}{2}\right) 
= \frac{2 R}{\sqrt{4 R^2 + (\epsilon_+ - A)^2}}$,
and $\cos\left(\frac{\theta_0}{2}\right) = \sqrt{\frac{1 +\langle \sigma_z (0) \rangle}{ 2}}$, $\epsilon_+ = \sqrt{A^2 +4R^2}$.
\begin{figure}[ht]
\subfigure[]{\includegraphics[width=4.0cm]{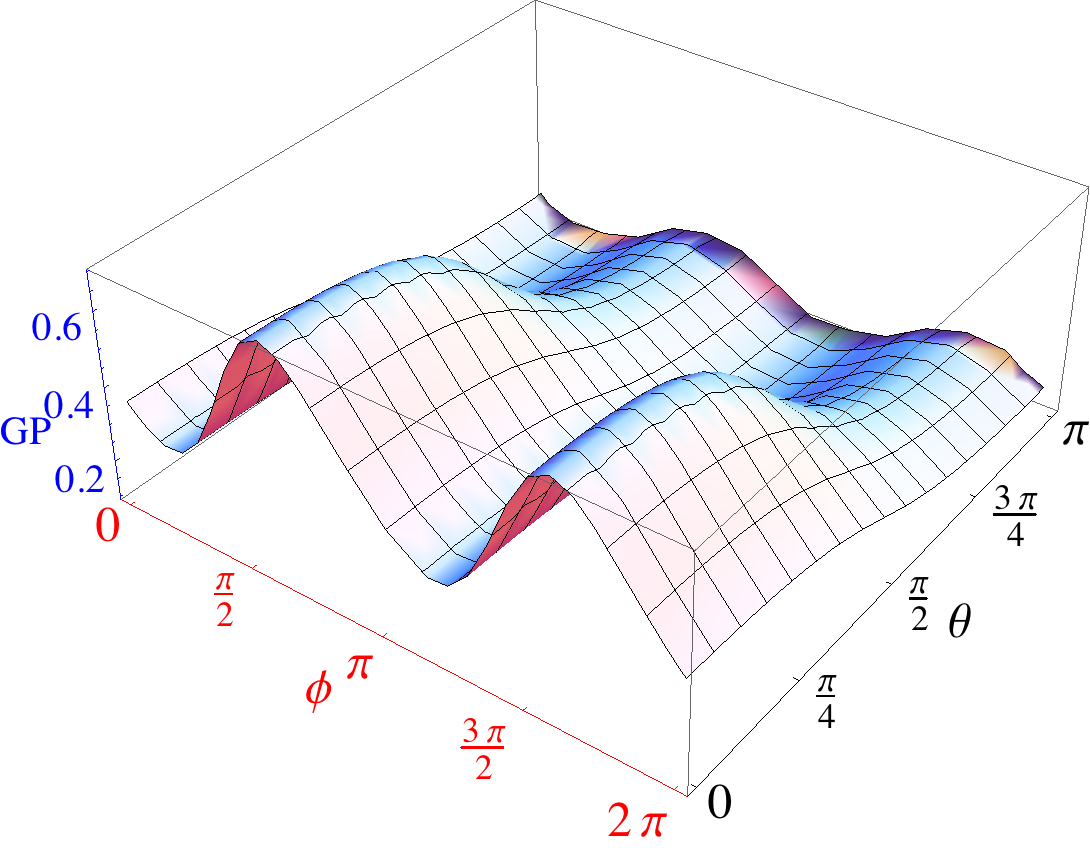}
\label{1a}}
\subfigure[]{\includegraphics[width=4.0cm]{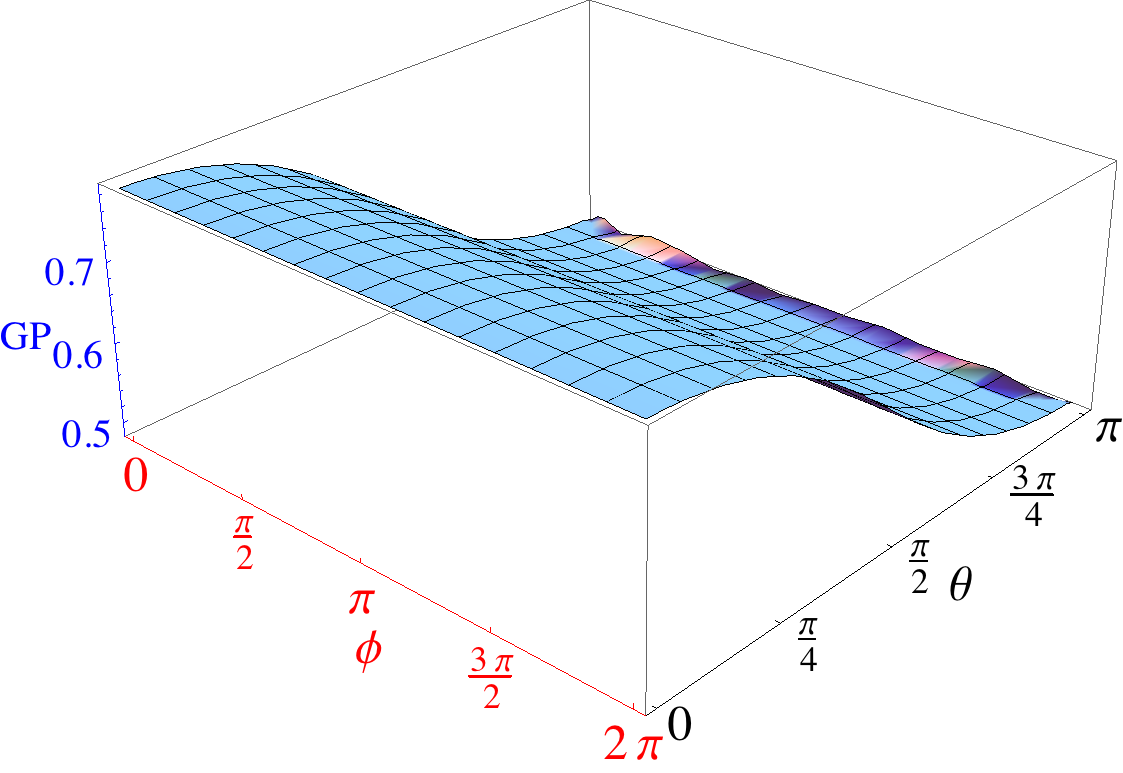}
\label{1b}}\\
\subfigure[]{\includegraphics[width=4.0cm]{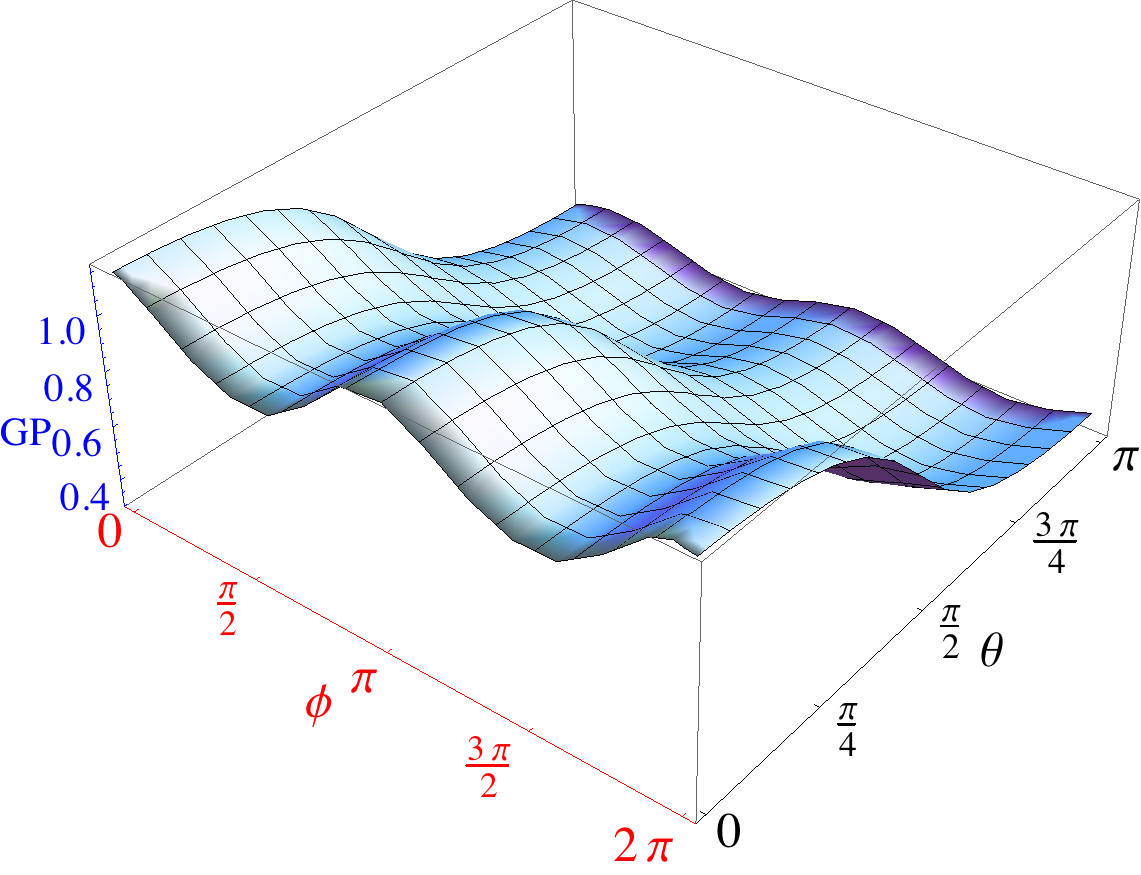}
\label{1c}}
\subfigure[]{\includegraphics[width=4.0cm]{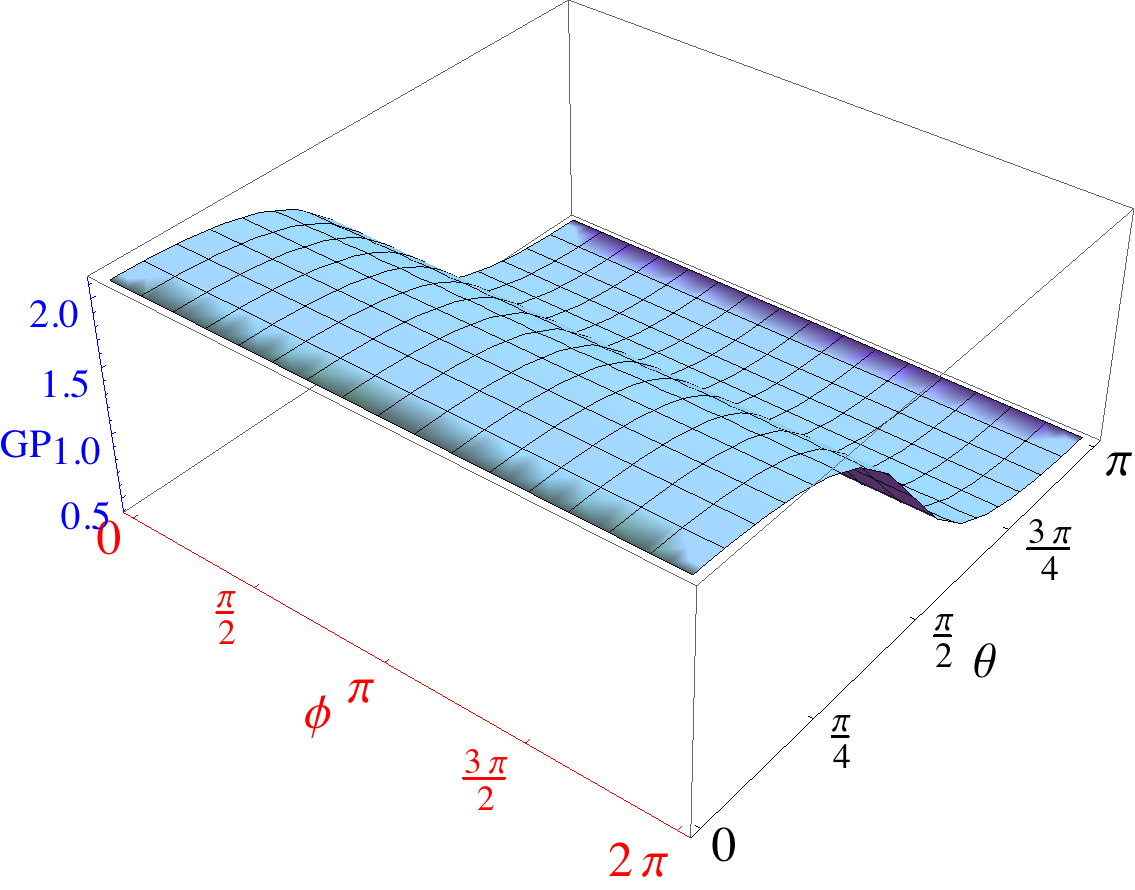}
\label{1d}}
\caption{(Color online) GP [$\gamma_g(\tau)$] with respect to $\theta$ and $\phi$ (a) when $\alpha_1 = 1$ and $\alpha_2 = 0$, that is, for the case of a single bath,  
(b) when  $\alpha_1 = \alpha_2 = \frac{1}{\sqrt{2}}$, (c) when  $\alpha_1 =\frac{\sqrt{3}}{2}$ and $\alpha_2 = \frac{1}{2}$ (d) when $\alpha_1 =\alpha_2 = \frac{1}{4}$. Here $\omega= 2$, and time $t=50$. A comparison 
between (a), (b), (c), and (d)  reiterates the point that the decay of GP gets frustrated when both the baths are acting and one of the best strategies is seen to be the case 
where $\alpha_1 = \alpha_2 = \frac{1}{4}$.}
\label{fig:1}
\end{figure}
The GP in the presence of two competing decoherence processes [Eq.\,\ref{4v}] can also be expressed as
\begin{eqnarray}
\gamma_g(\tau) &=& 
\tan^{-1}  \left[ 
\frac{ \sin  \frac{\theta_0}{ 2}  \cos  \frac{\theta_{\tau}}{ 2} \sin \delta \chi(t)  }{\cos \left( \frac{\theta_0}{ 2}  \right )
\sin  \frac{\theta_{\tau} }{ 2} + \sin \frac{\theta_0 }{ 2} \cos \frac{\theta_{\tau}}{ 2} \cos \delta \chi(t) } \right ]  \nonumber\\
&- & \int_0^{\tau} dt ~~\dot{\chi}(t) 
\cos^2 \frac{\theta_t}{ 2}, \label{4va}
\end{eqnarray}
where $\delta \chi(t) = (\chi(t) -\chi(0))$.
It can be easily seen from Eq.\,(\ref{4va}) that if we remove the influence
of the environment, we obtain for $\tau = \frac{2 \pi}{\omega}$, $\gamma_g =
-\pi(1- \cos(\theta_0))$, as expected,
which is the standard result 
for the unitary evolution of an initial pure state. 
If we take the angle $\theta_0 = \pi$, that is, the South Pole of the Bloch sphere of the spin of interest, 
then Eq.\,(\ref{4va}) simplifies to
\begin{eqnarray}
\gamma_g(\tau) =  \frac{1}{2} \int_0^{\tau} dt ~~\dot{\chi}(t) 
\left(1- \cos \theta_t \right). \label{GPsimple}
\end{eqnarray}
It can be noticed that a contribution to the GP, in Eq. (\ref{GPsimple}), coming from
the argument of the exponential, resembles the solid-angle expression for GP in the usual 
demonstrations.  

\begin{figure}[ht]
\subfigure[]{\includegraphics[width=4.25cm]{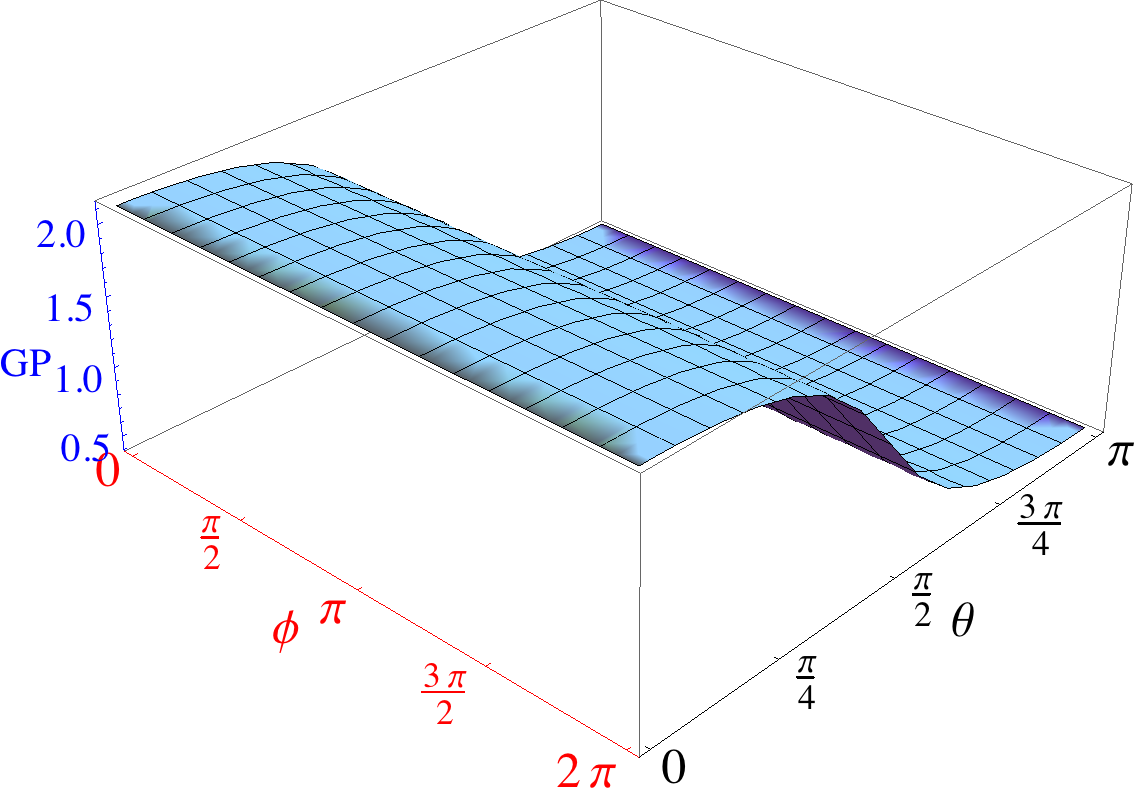}}
\subfigure[]{\includegraphics[width=4.25cm]{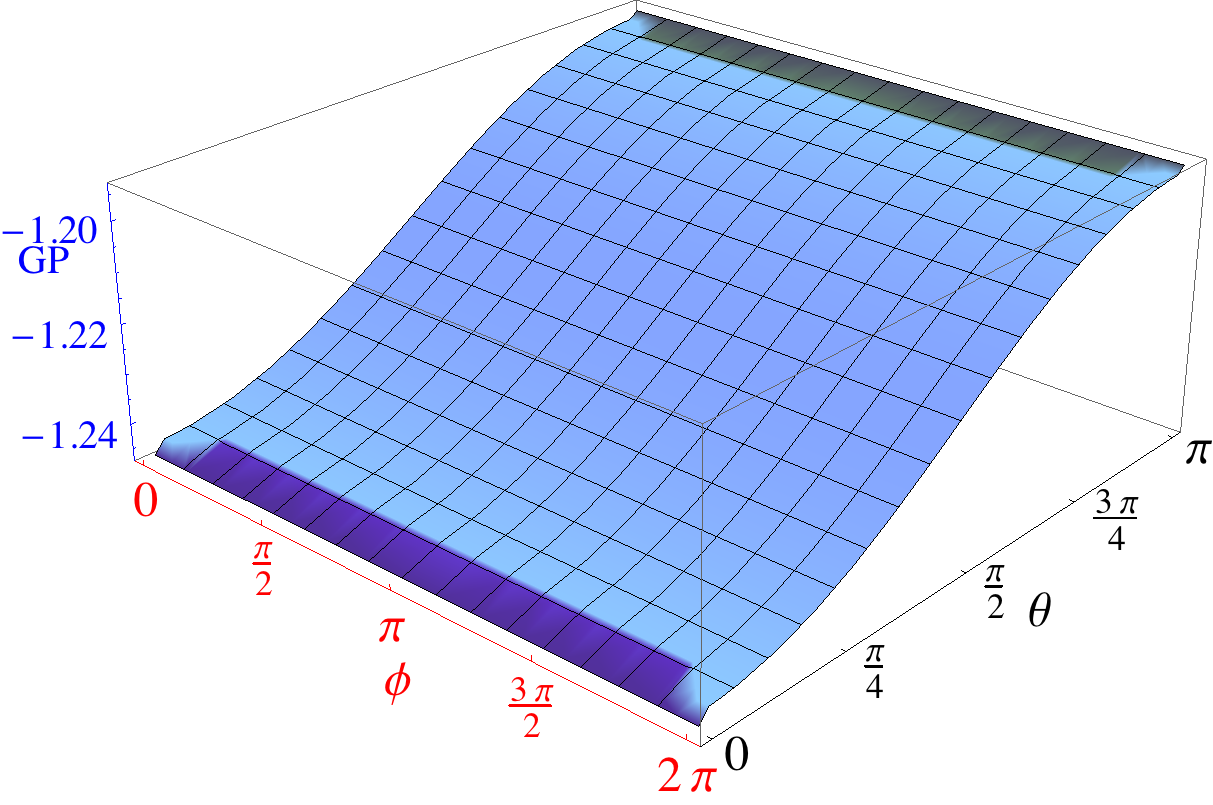}}
\caption{(Color online) GP for  $\alpha_1 = \alpha_2 = 1/4$ with respect to $\theta$ and $\phi$ for  (a) time $t = 50$ and (b) time $t = 200$, respectively.  Here $\omega= 2$. 
Among other sets of $\alpha_1$, $\alpha_2$, the figure corresponding to the case (a) is found to be optimum in resisting the depletion of GP. Therefore, for different time different $\alpha_1$ and $\alpha_2$ will help to enhance the GP.}
\label{fig:2}
\end{figure}

\begin{figure}[ht]
\includegraphics[width=8.0cm]{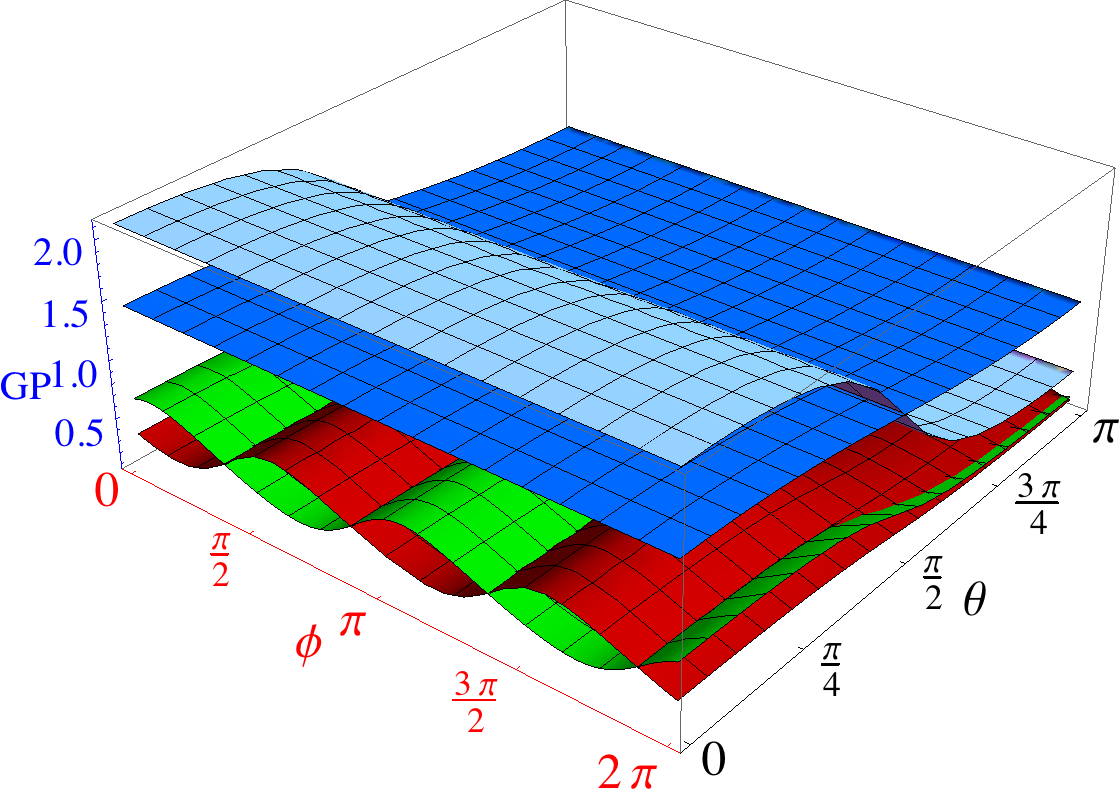}
\caption{(Color online) GP for  different $\alpha_1, \alpha_2$ with respect to $\theta$ and $\phi$ for time $t = 50$ and $\omega= 2$. 
The red curve corresponds to $\alpha_1 = 1$, $\alpha_2 = 0$; the green curve corresponds to $\alpha_1 = 0$, $\alpha_2 = 1$; the light blue curve to $\alpha_1 =\alpha_2=\frac{1}{4}$;
while the dark blue curve  corresponds to $\alpha_1 = \alpha_2 = \frac{1}{2}$. It is clearly evident from the plots that the decay of GP gets frustrated due to the 
presence of both the couplings $\alpha_1$ and $\alpha_2$.}
\label{fig:3}
\end{figure}

In Fig.\,\ref{fig:1}, GP  with respect to $\theta$ and $\phi$ for different values of coupling constants $\alpha_1$ and $\alpha_2$ for an evolution time $t=50$ and 
$\omega=2$ is depicted. A comparison between Figs.\,\ref{1a}, \ref{1b}, \ref{1c} and \ref{1d} where $\alpha_1 = 1$ and $\alpha_2 = 0$, $\alpha_1 = \alpha_2 = 1/\sqrt{2}$, $\alpha_1 = \sqrt{3}/2$ and $\alpha_2 = 1/2$, and 
$\alpha_1 = \alpha_2 = 1/4$, respectively, brings out the point that the decay of GP gets frustrated when both the baths are acting and one of the best strategy seems to be the case of $\alpha_1 =\alpha_2 = \frac{1}{4}$. In Fig.\ref{fig:2}, GP for $\alpha_1 = \alpha_2 = 1/4$ when $t=50$ and $t=200$ is shown and we can note that the optimum value of  $\alpha_1$ and $\alpha_2$ for 
maximum frustration of GP varies with time.  In Fig.\,\ref{fig:3}, a comparison is made of GP for  different coupling constants  with respect to $\theta$ and $\phi$ for 
time $t = 50$. For the QFD regime, that is, when $\alpha_1 \neq 0$ and $\alpha_2 \neq 0$ we observe that the value of GP is higher as compared to the case where QFD is not 
applicable, that is, for the case of single coupling constant. These observations bring out the inherent robustness of GP against decay of quantum fluctuations in the 
presence of QFD.

The problem of quantum frustration studied here, using the Hamiltonian, Eq.\, (\ref{eq:1}), is very general. 
These models can be understood by the fact that the two baths behave like  Goldstone modes, resulting from the spontaneous breaking of symmetry as is evident from the coupling to the baths by two noncommuting operators, 
such that the residual unbroken symmetry rotates the two Goldstone modes into each other. This is perfect when the two couplings are equal, but exists even for unequal couplings, a fact proved generally using renormalization group 
arguments in \cite{novais}. From the flow diagram of the two couplings, it is evident that the spin would remain coherent, irrespective of the strength of the spin coupling to the environment.  Spin coherence implies frustration of 
decoherence or the process of decoherence getting checked. Frustration of decoherence, in the present context implies that the decay of the off-diagonal terms in the density matrix of Eq.\, (\ref{reduceddm}) is reduced. This directly 
effects the terms $R$, $\chi$, and $\theta_t$ in Eqs.\, (\ref{4v}), (\ref{4va}) leading to an enhancement of GP, as compared to the 
case of coupling to a single bath or coupling to baths via commuting operators, that is, in the scenario of absence of frustration of decoherence.

\section{Analogy with Parrondo games}
\label{analogy}

The effect of  frustration on GP could be thought of as a Parrondo's game:
each game on its own is ``a single qubit interacting with its bath; one with $\sigma_x$,
and  with another  $\sigma_y$";
this would result in decoherence and dissipation leading to inhibition of GP. This would be the situation where each player looses his game. 
 However, when the two games are played in a synchronized fashion;
corresponding, here, to the case of ``the qubit interacting with two independent baths via non-commuting operators with coupling 
strengths $\alpha_1$ and $\alpha_2$",
then the decoherence and dissipation can get frustrated leading to improvement in GP over some range of parameters. Though, we have presented
the Parrondo like effect for GP for our model system, we expect this to be a generic feature of a quantum system interacting with two 
competing environments.

\section{Conclusions}
\label{conc}

 To conclude, by analyzing a simple model of QFD, we have illustrated the enhancement of geometric phase 
in the presence of two competing environments. The model being simple allows for an explicit evaluation, but is generic in the sense that it captures the essence of
frustration on GP for other models as well. Here we consider a qubit interacting with two independent baths via non-commuting operators, for e.g., $\sigma_x$, $\sigma_y$.
Naively, one would expect that due to interaction with two baths, the decoherence effect would increase leading to inhibition of geometric phase. 
However, in contrast to this, it is found that
decoherence gets suppressed: thus providing a typical framework for the Parrondo kind of game. Parrondo's games take place when a symmetry in the original problem gets 
broken. In this case the broken symmetry would be the interaction of the
qubit with the two independent baths via two non-commuting operators. Here a purely dephasing scheme would not work as that would require the system and interaction Hamiltonians 
to commute \cite{bg07}. But it is the non-commutativity of operators in the interaction Hamiltonian that leads to the Parrondo like effect for the geometric phase. 
This suggests that for quantum frustration of decoherence to be effective, we need both decoherence as well as dissipation. We hope that the effect found here 
can be used in fault tolerant quantum computation. This may also find wide applications in enhancement of geometric phases in other systems under competing decoherence.


\end{document}